\documentstyle[aps,preprint]{revtex}
\scrollmode
\begin{document}
\thispagestyle{empty}

\vspace{4cm}

\title{Sample-size dependence of the
ground-state energy in a one-dimensional localization problem.}

\vspace{4cm}

\author{C.Monthus$^{1}$, G.Oshanin$^{2,*}$,
 A.Comtet$^2$,  and S.F.Burlatsky$^{3}$}

\vspace{4cm}

\address{$^1$ Service de Physique Th\'eorique de Saclay, 91191 Gif-sur-Yvette
Cedex,
France\\
$^2$ Division de Physique Th\'eorique, Institut de Physique Nucl\'eaire, 91406
Orsay
Cedex, France\\
$^3$ Department of Chemistry BG - 10,
 University of Washington,
 Seattle, WA 98195 USA}

\vspace{4cm}

\address{\rm ( \today )}
\address{\mbox{ }}

\maketitle

\begin{abstract}

{We study the sample-size dependence of the ground-state
energy in a one-dimensional localization problem,
based on a supersymmetric quantum
mechanical Hamiltonian with random Gaussian potential.
We determine, in the form of bounds, the
precise form of this dependence and show that the disorder-average ground-state
energy
 decreases with an increase of the
size $R$ of the sample as a
stretched-exponential function,  $\exp( - R^{z})$,
where the characteristic exponent
$z$ depends merely on the nature of correlations in the random potential.
In the particular
case where the potential is distributed as a Gaussian white noise we prove that
$
z = 1/3$. We also predict the value of $z$ in
the general case of Gaussian random potentials with correlations.}
\end{abstract}
\address{\mbox{ }}

\address{\parbox{14cm}{\vspace{2cm}
}}


\vspace{2cm}


\pagebreak
\setcounter{page}{1}

\section{Introduction}

The one-dimensional Schr\"odinger Hamiltonian

\[\hat{H} \; = \; - \; \frac{d^{2}}{d x^{2}} \; + \; \phi^{2}(x) \; + \;
\frac{d
\phi(x)}{dx}  \quad (1)\]
arises in such diverse areas of quantum mechanics as studies of
 solitons in
conjugated polymers (polyacetylene) \cite{su,jack}, in which $\phi(x)$
describes
the dimerization pattern of the carbon-hydrogen sequence,  or of electrons
dynamics  in two-dimensional
systems subjected to random  magnetic fields \cite{coma}.
It also represents  a celebrated
toy model of
supersymmetric quantum mechanics introduced by Witten \cite{wit}.

In case of non-random potentials $\phi(x)$ the properties
of the Hamiltonian in Eq.(1)
have been thoroughly studied within the last two decades
and several interesting results
have been obtained.
In particular, exact solutions of the Schr\"odinger equation
 have been derived for a class of the so-called shape
invariant potentials \cite{dutt}. Besides, such form of the
 Hamiltonian  inspired a new method
of semi-classical quantization \cite{comb,ino}.

More recent studies of $\hat H$ have focused on situations in
which $\phi(x)$ is random \cite{coma,bouc,bouca,comc,cec}.
Here Eq.(1) describes a localization problem in which, remarkably, the average
density
of states, the localization length and the Lyapunov exponent can be computed
exactly
\cite{bouc,bouca,comc}. In fact, this problem represents an additional
to the two known by now
\cite{and,llo,tou} examples of localization problems which can be solved
exactly
 in
continuum.

In the present paper we study a new aspect of the one-dimensional
localization problem associated with the Hamiltonian in Eq.(1) in which the
potential $\phi(x)$ is a random function of the space variable $x$.
 We focus on the physically interesting situation  where
$\hat H$ is defined on a finite interval $[ - R, R]$ and
analyse the  sample-size dependence of the ground-state
 energy $E_{0}(R,\{\phi\}) $, which may be expressed in terms of non-local
exponential
 functionals of random potential $\phi(x)$. We consider mostly throughout the
paper
the case where $\phi(x)$ is Gaussian, $\delta$-correlated white
noise with the moments

\[<\phi(x)> \; = \; 0 \quad (2)\]
and
\[<\phi(x) \phi(x')> \; = \; \sigma \delta(x - x'), \quad (3)\]
where the brackets denote the
average with respect to realizations of $\phi(x)$.
The relevance of short-range correlations
in the distribution of $\phi(x)$ and their influence on the
sample-size dependence
are also succinctly addressed and several
predictions are made.
We find that for such potentials the disorder-averaged ground state (DAGS)
energy
$$
E_{0}(R) \equiv < E_{0}(R,\{\phi\}) >
$$
decreases as the size $R$ grows
as a stretched-exponential function, $E_{0}(R)
\propto \exp( - R^{z})$, where the characteristic exponent $z$ depends only on
 the
correlation properties of the random potential. In the particular case where
fluctuations in $\phi(x)$ are delta-correlated, as in Eq.(3), we prove that $z
=
 1/3$. The value of the exponent $z$ is also predicted, on heuristic grounds,
for the general case of random Gaussian potentials with correlations.
We show that such a stretched-exponential dependence on $R$ stems
from atypical realizations of $\phi(x)$, which are reminiscent of the
representative
trajectories supporting long-time anomalous decay laws of
 the survival
probability for diffusion in the presence of randomly placed traps
\cite{bal,don}
or of the Lifschitz
 singularities \cite{lif} in the low-energy spectrum of an electron in the
prese
nce
of randomly dispersed scatterers.
Our results are presented in form of lower and upper bounds on $E_{0}(R)$,
which show the same dependence on $R$ and thus define this dependence exactly.
 The method of the derivation of bounds, which we invoke here,  has been
previously discussed
in \cite{bura,oshb} and is based on the statistics of extremes of random
potential
 $\phi(x)$.

The paper is structured as follows: In Section 2 we present a simple
derivation of  the ground-state energy $E_{0}(R,\{\phi\})$ in case of
deterministic potentials $\phi(x)$. We recover in
a straightforward way an explicit formula
for $E_{0}(R,\{\phi\})$, which coincides when the appropriate notations are
introduced
with the one found in \cite{bart}, where the influence of the finite-size
effects
on the ground-state energy in conventional quantum mechanics has been examined.
We then present some arguments showing that this formula is still valid
in the case of a random potential
 $\phi(x)$. In Section 3 we estimate
the sample-size dependence of the ground-state energy considering only
 typical realizations
of random potential $\phi(x)$. Employing the standard Jensen inequality we show
then
that such an estimate constitutes a lower bound on the average ground-state
energy.
Additionally,
we illustrate that such an estimate
allows to recover the correct low-energy behavior
 of the integrated density of states of
Hamiltonian (1).
Further on, in Section 4, we devise a more accurate approach
 and derive a lower bound (subsection $A$) and an upper bound (subsection $B$)
on
$E_{0}(R)$, which determine exactly its behavior in the large-$R$ limit. In
subsection
$C$ we address the question of the DAGS behavior in situations, in which the
fluctuations of the random potential $\phi(x)$ are correlated, and also discuss
some similar features between
 the realizations of disorder supporting large-$R$ behavior
 of the DAGS and the realizations of random walks
 which support anomalous long-time tails
 of the survival probability  for the diffusion in the presence of traps.
Finally, in Section 5 we conclude with a summary of our results.

\section{Calculation of the energy shift in a finite sample.}

The structure of the Hamiltonian in Eq.(1) is such that, for an arbitrary
 function $\phi(x)$, two independent solutions of the Schr\"odinger equation

\[\hat{H} \varphi_{0}^{(1,2)}(x) \; = \; 0, \quad (4)\]
may be explicitly expressed as functionals of $\phi(x)$ :

\[\varphi_{0}^{(1)}(x) \; = \; \exp(\int^{x}_{0} dx' \phi(x'))    \quad (5)\]
and
\[\varphi_{0}^{(2)}(x) \; =
\; \varphi_{0}^{(1)}(x) \; \int^{x}_{0}  \frac{dx}{[\varphi_{0}^{(1)}(x)]^{2}}
     \quad (6)\]

\subsection{Deterministic potentials $\phi(x)$.}

In this subsection we consider the case of deterministic potentials
 $\phi(x)$, which
decay fast enough when $x \to \pm \infty$ to make $\varphi_{0}^{(1)}$
normalizable on the whole axis, i.e.
$$
\int_{-\infty}^{\infty} dx \big[ \varphi_{0}^{(1)}(x) \big]^2 < \infty
$$
Therefore,   $\varphi_{0}^{(1)}$ is  a zero-energy
wave-function of the Hamiltonian $\hat{H}$ defined on the whole axis.

Consider now how the situation will be changed if one
assumes that the Schr\"odinger equation with the Hamiltonian in Eq.(1)
is defined not on the whole $X$-axis, but only on a finite interval $[-R, R]$.
The new ground-state wave-function $\hat{\Psi}_{0}(x)$ on this finite interval
satisfies the Schr\"odinger equation

\[[  - \; \frac{d^{2}}{d x^{2}} \; + \; \phi^{2}(x) \; + \; \frac{d
\phi(x)}{dx}] \; \hat{\Psi}_{0}(x) \; = \; E_{0}(R,\{\phi\}) \;
\hat{\Psi}_{0}(x
),
\quad (7)\]
with an a priori unknown energy $E_{0}(R,\{\phi\})$ that will depend on
the explicit form of the potential $\phi(x)$.
Eq.(7)
has to be supplemented by the following Dirichlet boundary conditions
at the ends of the interval, i.e. at points $x = - R$ and $x = R$,

\[ \hat{\Psi}_{0}(x = - R) \; = \; \hat{\Psi}_{0}(x = R) \; = \; 0  \quad (8)\]
Our goal will be to estimate the energy shift $E_{0}(R,\{\phi\})$ of the ground
state
caused by the introduction of boundary conditions imposed on a finite interval.

Multiplying both sides of Eq.(7) by $\varphi_{0}^{(1)}(x)$ and
integrating from $-R$ to $R$,
we get the following identity

\[E_{0}(R,\{\phi\}) \int^{R}_{-R} dx \; \varphi_{0}^{(1)}(x)
\; \hat{\Psi}_{0}(x) \; = \quad\]
\[ \; = \; \int^{R}_{- R} dx \; \varphi_{0}^{(1)}(x) \; [  -
\; \frac{d^{2}}{d x^{2}} \; + \; \phi^{2}(x) \; + \; \frac{d
\phi(x)}{dx}] \; \hat{\Psi}_{0}(x) \quad (9)\]
Integrating by parts the kinetic term on the right-hand-side of Eq.(9) yields

\[E_{0}(R,\{\phi\}) \int^{R}_{- R} dx \; \varphi_{0}^{(1)}(x) \;
\hat{\Psi}_{0}(
x)
\; = \; \quad\]
\[ = \; \varphi_{0}^{(1)}(- R) \; \frac{d \hat{\Psi}_{0}(x)}{dx}_{|x = - R} \;
-
 \;
\varphi_{0}^{(1)}(R) \; \frac{d \hat{\Psi}_{0}(x)}{dx}_{|x = R}    \quad (10)\]

To proceed further on, we assume that $R$ is sufficiently large and
estimate the behavior of the derivative of
$\hat{\Psi}_{0}(x)$ in this limit. Since $E_{0}(R,\{\phi\})$
vanishes when $R \rightarrow \infty$, one  expects that the ground-state
wave-function on the finite (large) interval
  may be well approximated by a suitable linear combination of
the two independent zero-energy solutions,  $\varphi_{0}^{(1)}(x)$ and
$\varphi_{0}^{(2)}(x)$. In the vicinity of $x = R$ one has

\[\hat{\Psi}_{0}(x) \; \approx \; \varphi_{0}^{(1)}(x) \; [ 1 \; + \; \alpha \;
\int^{x}_{0}  \frac{dx}{[\varphi_{0}^{(1)}(x)]^{2}}] \quad (11)\]
where $\alpha$ is determined through the boundary condition at $x = R$, which
gives

\[\alpha \; = \; -  \; \{\int^{R}_{0}
\frac{dx}{[\varphi_{0}^{(1)}(x)]^{2}}\}^{
 -1},
 \quad (12)\]
and consequently, we find that the derivative of $\hat{\Psi}_{0}(x)$ obeys

\[\frac{d \hat{\Psi}_{0}(x)}{dx}_{|x = R} \;
\approx \; \frac{\alpha}{\varphi_{0}^{(1)}(R)} \;
= \; - \; \{\varphi_{0}^{(1)}(R) \; \int^{R}_{0}
\frac{dx}{[\varphi_{0}^{(1)}(x)]^{2}}\}^{-1} \quad
(13)\]

Similarly, for  the derivative of the ground-state wave-function
in the vicinity of $x = - R$ one finds

\[\frac{d \hat{\Psi}_{0}(x)}{dx}_{|x = - R} \; \approx \; - \;
\{\varphi_{0}^{(1)}( - R) \; \int^{ - R}_{0}
\frac{dx}{[\varphi_{0}^{(1)}(x)]^{2}}\}^{-1} \quad
(14)\]

Now, combining Eqs.(10) to (14) we have

\[E_{0}(R,\{\phi\}) \int^{R}_{- R} dx \; \varphi_{0}^{(1)}(x) \;
\hat{\Psi}_{0}(
x)
\; =
\; \{\int^{0}_{-R}  \frac{dx}{[\varphi_{0}^{(1)}(x)]^{2}}\}^{-1} \; +
\; \{\int^{R}_{0}  \frac{dx}{[\varphi_{0}^{(1)}(x)]^{2}}\}^{-1} \quad (15)\]
and, finally, replacing in the integral on the left-hand-side of
Eq.(15) the function $\hat{\Psi}_{0}(x)$ by $\varphi_{0}^{(1)}(x)$ (which
results
in exponentially small with $R$ errors \cite{bart}) we find the following
explicit
expression for the
ground-state energy on a finite interval

\[E_{0}(R,\{\phi\}) \; \approx \;
\frac{1}{\int^{0}_{-R} [\varphi_{0}^{(1)}(x)]^{- 2} \; dx \;
\int^{R}_{- R} [\varphi_{0}^{(1)}(x')]^{2} \; dx'} \; +  \quad\]
\[ \; + \;  \frac{1}{\int^{R}_{0} [\varphi_{0}^{(1)}(x)]^{- 2} \; dx \;
\int^{R}_{- R} [\varphi_{0}^{(1)}(x')]^{2} \; dx'} \;    \quad (16)\]

Equation (16) reproduces the result of \cite{bart}, derived in terms of a
different
approach, when the notation $V(x) =
\phi^{2}(x) + d \phi(x)/dx$ is introduced.
We note also parenthetically that
 expressions of
quite identical structure  were recently obtained for the diffusion constant
for
random motion in an external periodic potential \cite{alm,leba,lebb,ruck}.
Therefore, the results which will be
obtained in the following also apply to this problem, provided that the
potential
 is defined as in Eqs.(2) and (3).

\subsection{Random potentials $\phi(x)$.}

We now have to explain why Eq.(16) is still valid in the case of random
potentials
 $\phi(x)$. We first mention that the analysis of the previous subsection is
based on
the assumption that the wave-functions are normalizable on the whole line. In
case
of random potentials $\phi(x)$, as defined
in Eqs.(2) and (3), the integral
$\int^{R}_{0} dx \phi(x)$ may show an unbounded
growth when $R \rightarrow \infty$ and thus
the wave-functions are not normalizable.
 Therefore, it is not possible
 to use directly the
results of the previous subsection. One may, however, notice that
 due to the presence of disorder the zero-energy
solutions are localized. This observation allows us to get rid off the "bad"
behavior
at infinity. For a given realization of random potential $\phi(x)$ one chooses
$R$ in
such a way that

\[\phi(-R) \; > \; 0   \; \text{and} \; \phi(R) \; < \; 0   \quad (17)\]
and then define an auxiliary configuration $\tilde{\phi}(x)$ (see Figs.1a and
1b)
 such that

\[\tilde{\phi}(x) \; = \; \phi(x)  \; \text{for} \; - R \; < \; x \; < \; R,
\quad (18)\]
\[\tilde{\phi}(x) \; = \; \phi( - R)  \; \text{for} \;  x \; < \; - \; R, \quad
(19)\]
and
\[\tilde{\phi}(x) \; = \; \phi(R)  \; \text{for} \;  x \; > \;  R, \quad (20)\]
The corresponding wave-function $\tilde{\Psi}_{0}(x) \; = \; \exp( \int^{x}_{0}
\tilde{\phi}(x') dx')$ is therefore an exact zero mode of the Hamiltonian

\[\tilde{H} \; = \; \; - \; \frac{d^{2}}{d x^{2}} \; + \; \tilde{\phi}^{2}(x)
\;
 +
\; \frac{d
\tilde{\phi}(x)}{dx}  \quad (21)\]
on the whole line. When $R$ is sufficiently large, it follows from our previous
discussion that there exists a ground-state wave-function
 of the Hamiltonian $\tilde{H}$ on the interval $[ - R, R]$, whose ground-state
energy
is given by Eq.(16). Since this state has its support on $[ - R, R]$ the
functions
$\tilde{\Psi}_{0}(x)$ and $\varphi_{0}(x)$ coincide, and thus it is also a
quasi zero
 mode of
$\hat{H}$ with the ground-state energy defined by  Eq.(16).

\section{Typical realizations of  disorder and the Jensen inequality.}

We start our analysis of the behavior of the disorder-average ground-state
energy
in Eq.(16)  considering first the typical realizations
of the random potential $\phi(x)$. To do this, we rewrite
the ground-state energy for a given sample
with a particular realization of the random potential $\phi(x)$, Eq.(16),
in terms of the following exponential functionals of the potential $\phi(x)$:

\[\tau_{\pm}(z,y) \; = \; \int^{y}_{z} dx \; [\varphi_{0}^{(1)}(x)]^{\pm 2} \;
=
\; \int^{y}_{z} dx \; \exp( \pm 2 \int^{x}_{0} \phi(x') dx'), \quad (22)\]
which gives

\[ E_{0}(R, \{\phi\}) \; = \; {1 \over {\tau_{+}(-R,R)}}
 \bigg({1 \over {\tau_{-}(-R,0)}} + {1 \over { \tau_{-}(0,R)} } \bigg)
\quad (23)\]

We notice now that the function  $W(x) =  \int^{x}_{0} dx \phi(x)$,
which appears in the
definition of $\tau_{-}(0,R)$, Eq.(22), for the potentials
 as in Eqs.(2) and (3) is simply a
trajectory of a symmetric random walk. Consequently,
for typical realizations of the
random potential $\phi(x)$ the value of $W(R)$ should be of order
$  (\sigma R)^{1/2}$; hence
 $\tau_{-}(0,R)$ should grow typically as $\exp( 2 (\sigma R)^{1/2})$ and then
Eq.(23)
 entails
$ E_{0}(R)  \propto \exp( - 4 (\sigma R)^{1/2})$. Therefore, for
most realizations of $\phi(x)$ one may expect that $E_{0}(R, \{\phi\})$
vanishes
 with an increase of $R$ as
$$
E_{0}(R, \{\phi\})_{typ}  \sim \exp( - \sigma^{1/2} R^{1/2})
\quad (24)
$$

This typical behavior may be used to evaluate correctly the
low-energy asymptotic behavior of the density of states of the Hamiltonian in
Eq.(1).
Eq.(24) means that a wave-function of low energy $E$ has
 typically a spatial extension $2 R$ such that

\[ R \; \propto \frac{ \ln^{2}( E)}{\sigma}  \quad (25)\]
Therefore, the number of such states per unit length behaves typically as

\[N(E) \; \propto \; \frac{1}{2 R} \; \propto \; \frac{\sigma}{ \ln^{2}(E)},
\quad (26)\]
It is now interesting to compare Eq.(26) with the exact result \cite{bouca}:

\[N(E) \; = \frac{2 \sigma}{\pi^{2}} \; \frac{1}{ J^{2}_{0}(z) \; + \;
N_{0}^{2}
(z)},
\quad (27)\]
where $z = \sqrt{E}/\sigma$ and $J_{0}, N_{0}$ are Bessel functions. In the
limit
 $E\rightarrow 0$ one has from Eq.(27) that $N(E)  \sim
 2 \sigma ln^{-2}(E/4 \sigma^{2})$, i.e. the behavior which is quite consistent
with
our estimate in Eq.(26).  Therefore, Eq.(26) shows that
 anomalous singular
behavior of the
integrated density of states is supported by typical realizations of disorder
and thus
is quite
distinct from the Lifschitz singularity \cite{lif}, which
 is most often encountered in disordered quantum mechanical systems.

We now show that the estimate based on typical realizations of the disorder
represents actually
a lower bound on the DAGS energy. To show it explicitly we invoke
the standard Jensen
inequality between the average of the exponential of some function $F$ and the
exponent
of the averaged value of $F$,

\[< \exp[ - F]> \; \geq \; \exp[ - <  F >] \quad (28)\]
Now, choosing

\[F \; = \; - \ln  E_{0}(R, \{\phi\})
\quad (29)\]
we find, taking into account  Eq.(23),
that $E_{0}(R)$ is
bounded from below by

\[ E_{0}(R)  \; \geq \; \exp[ <ln(\tau_{-}(-R,R))> \; -
\; <ln(\tau_{+}(-R,R))>
\quad \]
\[ - \; <ln(\tau_{-}(-R,0))> \; - \; <ln(\tau_{-}(0,R))> ] \quad (30) \]

One may readily notice that for any random function $\phi(x)$ of zero mean,
 not all terms in the exponent on the right-hand-side of Eq.(30) are to
be calculated independently; obviously,

\[ <ln(\tau_{-}(-R,R))> \; = \; <ln(\tau_{+}(-R,R))>  \quad (31)\]
and

\[ <ln(\tau_{-}(-R,0))> \; = \; <ln(\tau_{-}(0,R))>  \quad (32)\]
Consequently,
the first two terms in the rhs of Eq.(30) cancel each
other and we have only to perform
averaging of  $ln(\tau_{-}(0,R))$.

These functionals $\tau_{\pm}(0,R)$ appear in different physical
backgrounds  \cite{deC,der}. Their discrete-$x$ counterpart,
which is the sum of products of
independent random variables of the form

\[ \tau_{-}(N) \; = \; 1 \; + \; z_{1} \; + \;  z_{1} z_{2} \; +
\;  z_{1} z_{2} z_{3} \; + \; ... \;
+ \; z_{1} z_{2} z_{3} ... z_{N}, \; \quad\]
with
\[ \; z_{n} \; = \; \exp(\phi_{n}),  \quad\]
is known as the Kesten variable \cite{kes} and plays an
important role in the theory of renewal
processes. The distribution function  of the continuous-$x$ functional
$\tau_{-}(0,R)$
has been recently
examined in \cite{bura,osha,mon,oshb}
within the context of diffusion in the presence of a random quenched force
(the Sinai diffusion \cite{sin,bouca})
and also in the literature on mathematical finance \cite{yor}.

The average
logarithm of the functional $\tau_{-}(0,R)$ can be obtained from  the
 probability distribution of this functional \cite{osha,mon,oshb}:

\[< ln(\tau_{-}(0,R))> \; =
\; {2 \over \pi} \int_0^{\infty} {dk \over k^2} \left( 1- \exp(-2 \sigma R k^2)
\pi k \coth (\pi k) \right)  - \Gamma'(1)
-\ln(2 \sigma) \quad\]
\[ \approx \; \left(\frac{8 \sigma R}{\pi}\right)^{1/2} \;
  - \Gamma'(1) -\ln(2 \sigma) \; + \; O(\frac{1}{\sigma R}),
\quad (33b)\]
where the notation
$O(1/R)$ means that the neglected terms multiplied by $R$
 will give a constant as $R \rightarrow \infty$. Eq.(33b) has been recently
rederived in
\cite{kree} which tested predictions of the replica variational approximation
\cite{mez} for a particular physical system - a classical particle in a
one-dimensional
box subjected to a random potential which constitutes a Wiener process on the
coordinate axis \cite{sin,bouca}. A detailed discussion of
the average
logarithm of the functional $\tau_{-}(0,R)$ can be found in \cite{CMY}.

Accordingly, for the DAGS energy we obtain

\[E_{0}(R)  \; \geq \; \exp( - 4 (\frac{2 \sigma R}{\pi})^{1/2}), \quad (34) \]
which thus shows that
 $E_{0}(R)$ vanishes with an increase of the sample size $R$ not faster than
 a
stretched-exponential function $\exp( - R^{z})$ with $z = 1/2$.
However, this lower bound, which is supported by typical realizations of
disorder
may be improved as we will see in the next section.

\section{Lower and upper bounds on $E_{0}(R)$.}

In this section we set out to show that,  in the
limit $R \rightarrow \infty$, the  dependence of
the disorder averaged ground state energy $E_{0}(R)$ on $R$
 is quite different from that in Eq.(34).
Here we will derive more
accurate bounds
which show that in the large-$R$ limit, the actual dependence of
the disorder averaged ground-state energy $E_{0}(R)$ on $R$ is described by
 a stretched-exponential
function  $\exp( - R^{z})$ but with a smaller
exponent, $z = 1/3$. This means that large-$R$ behavior of $E_{0}(R)$ is
 supported by atypical realizations of $\phi(x)$. These realizations will be
also
specified below.

\subsection{A lower bound.}

Let us begin with the derivation of a lower bound on $E_{0}(R)$. We first note
that
since $\phi(x)$ enters the expression for $E_{0}(R,
\{\phi\})$ only in the form $\int dx \phi(x)$,
averaging  with respect to realizations of $\phi(x)$ amounts actually to
 the averaging over different trajectories $W(x)$
of a symmetric random walk.
Therefore, we can formally write down
the average as a product of two  path-integrals

\[E_{0}(R) \; = \; < E_{0}(R,
\{\phi\}) > \; = \; \int_{\Omega} D\{W(x)\} \int_{\Omega'} D\{W'(x)\}
\; P[W(x)] \; P[W'(x)] \; E_{0}(R, \{\phi\}), \quad (35)\]
where the notations used have the following meaning: The symbol $\Omega$
denotes the set of  $all$
$possible$ (unrestricted) trajectories $W(x)$
of a symmetric random walk, which "starts" at $x = 0$ at the
origin $W(0) = 0$, and "time" variable $x$ is defined on the interval $[0, R]$.
We describe schematically the set $\Omega$  in Fig.2, where for
notational convenience we use the discrete-$x$ picture and depict it using
the axis $W(x)$ and $x$, i.e. using
"directed polymers"-like representation.
The trajectories $(1)$ and $(2)$ are two examples
of possible trajectories which belong to the set $\Omega$.
The symbol $\Omega'$ denotes, correspondingly, the
set of all possible trajectories $W'(x)$ with the "time" variable $x$ defined
on
 the
interval $[0, - R]$. The trajectories in $\Omega$ and $\Omega'$  are
statistically uncorrelated.
Finally,
the symbols $D\{W(x)\}$ and $D\{W'(x)\}$ denote that the integration is
performed
along the
trajectories $W(x)$ and $W'(x)$; $P[W(x)]$ (or $ P[W'(x)]$) is the
corresponding
 measure
of a given trajectory $W(x)$ (or $W'(x)$), which is
the standard Wiener measure.

The next essential step is as follows. Suppose that from the entire set
$\Omega$
(and $\Omega'$) we select some amount of trajectories having certain
 prescribed properties
and denote this subset of $\Omega$ ($\Omega'$) as $\omega$ ($\omega'$). Then,
for any
positive definite functional $E_{0}(R, \{\phi\})$ the following inequality
holds

\[\int_{\Omega} D\{W(x)\} \int_{\Omega'} D\{W'(x)\} \; P[W(x)] \; P[W'(x)] \;
\; E_{0}(R, \{\phi\}) \; \geq \; \quad\]
\[ \; \geq \; \int_{\omega} D\{W(x)\} \int_{\omega'}
D\{W(x)\} \; P[W(x)] \; P[W'(x)] \;
\; E_{0}(R, \{\phi\}), \quad (36)\]
where the integrations in the rhs of Eq.(36) extend only over the
trajectories which belong to the subsets $\omega$ and $\omega'$ of
the entire sets $\Omega$ and $\Omega'$.
Employing
the inequality in Eq.(36), we get the following bound

\[ E_{0}(R)  \; \geq \; \int_{\omega} D\{W(x)\} \int_{\omega'}
D\{W'(x)\} \; P[W(x)] \; P[W'(x)] \;
\; E_{0}(R, \{\phi\}) \quad (37)\]

Now we define the subset $\omega$ ($\omega'$) as follows (Fig.2):
$\omega$ ($\omega'$) is the set of all
 trajectories $W(x)$ ($W'(x)$), which, for any $x$ from
the interval $[0, R]$ ($[0, - R]$ for $\omega'$),
remain inside the strip $[- A, A]$,
 i.e. such
trajectories $W(x)$ (and $W'(x)$) which obey $- A \leq W(x),W'(x) \leq A$ for
any
$x$ from the interval $[0, R]$ ($[0, - R]$ for $\omega'$). In Fig.2
trajectories
 which
form the subsets $\omega$ and $\omega'$ are exemplified by $(2)$ and $(2')$.

Next, we diminish the rhs of Eq.(37), i.e. enhance  the inequality in Eq.(37),
by substituting instead of  $E_{0}(R,
\{\phi\})$ its minimal value on the subsets $\omega$ and $\omega'$. By
definition of
$\omega$ and $\omega'$, which implies that $|W(x)| \leq A$ and $|W'(x)| \leq A$
we have

\[\int^{R}_{- R} dx \exp( 2 W(x)) \; \leq \; 2 R \exp( 2 A), \quad (38)\]
\[\int^{0}_{-R} dx \exp( - 2 W(x)) \; \leq  \;  R \exp( 2 A), \; \quad (39.a)\]
and
\[\int^{R}_{0} dx \exp( - 2 W(x)) \; \leq  \;  R \exp( 2 A) \quad (39.b)\]
Consequently, for any realization of $W(x)$ or $W'(x)$ which belongs to the
subs
ets
 $\omega$
and $\omega'$, the following inequality holds

\[ E_{0}(R,
\{\phi\}) \; \geq \; min_{\omega,\omega'} \{E_{0}(R,
\{\phi\})\} \; = \; \frac{\exp( - 4 A)}{2 R^{2}}  \quad (40)\]
Substituting Eq.(40) into Eq.(37) we find the following  bound:

\[ E_{0}(R)
\; \geq \; \frac{\exp( - 4 A)}{2 R^{2}} \; \int_{\omega} D\{W(x)\}
\int_{\omega'}
D\{W'(x)\} \; P[W(x)] \; P[W'(x)] \;  \quad (41)\]
We notice now that the product of integrals along two "restricted"
(statistically
 independent)
paths $W(x)$ and $W'(x)$
on the rhs of Eq.(41) is equal to the
probability that two independent random walkers during "time" $R$ will remain
within
the strip $[- A, A]$, which means that

\[\int_{\omega} D\{W(x)\} \int_{\omega'}
D\{W'(x)\} \; P[W(x)] \; P[W'(x)]
 \; = \; P^{2}(A,R), \quad (42)\]
where $P(A,R)$ is the corresponding probability for a single random walker
\cite{weiss}:

\[P(A,R) \; = \; \frac{4}{\pi} \; \sum_{k = 0}^{\infty} \; \frac{(- 1)^{k}}{2 k
+ 1} \;
\exp( - \frac{( 2 k + 1)^{2} \pi^{2} \sigma R}{8 A^{2}})  \quad (43)\]

Combining Eq.(41) to (43) we thus obtain

\[E_{0}(R) \; \geq \; \frac{8 \;  \exp( - 4 A)}{\pi^{2} R^{2}} \;
(\sum_{k = 0}^{\infty} \; \frac{(- 1)^{k}}{2 k + 1} \;
\exp( - \frac{( 2 k + 1)^{2} \pi^{2} \sigma R}{8 A^{2}}))^{2} \quad (44)\]

The function in the rhs of Eq.(44) contains a free trial parameter $A$. The
inequality
in Eq.(44) holds for any value of this parameter and thus represents a family
of
 lower
bounds. Therefore, we will choose such a value of $A$, which maximizes the rhs
of
Eq.(44) and thus defines the maximal lower bound.
For $R$ sufficiently large the maximal contribution to the probability
distribution in
Eq.(43) comes from the term with $k = 0$, i.e.

\[P(A,R) \; \approx \; \frac{4}{\pi} \exp( - \frac{\pi^{2} \sigma R}{8 A^{2}}),
\quad (45)\]
and consequently,

\[E_{0}(R) \; \geq \; \frac{8}{\pi^{2} R^{2}} \; \exp( - 4 A - \frac{\pi^{2}
\sigma
R}{4 A^{2}}) \quad (46)\]
Taking the derivative of the rhs of Eq.(46) with respect to the parameter $A$,
we find
that

\[A \; = \; A^{*} \; = \; \frac{1}{2} (\pi^{2} \sigma R)^{1/3} \quad (47)\]
provides its maximal value. Substituting Eq.(47) into Eq.(46) we thus arrive at
the
following "maximal" lower bound

\[E_{0}(R) \; \geq
\; \frac{8}{\pi^{2} R^{2}} \; \exp( - 3 (\pi^{2} \sigma R)^{1/3}), \quad (48)\]
which shows that in the large-$R$ limit the DAGS energy
vanishes not faster than $\exp( - R^{1/3})$, i.e. at a slower rate that
"typical behavior" in
Eq.(34). This improved lower bound in Eq.(48) is supported by atypical
realizations of $W(x)$, such that $W(x) \propto x^{1/3}$, i.e. by trajectories
of
$W(x)$ which are spatially more
confined than "typical" realizations of random walk trajectories for which
$W(x)
\propto x^{1/2}$.

\subsection{An upper bound.}

Let us discuss now the derivation of an upper bound on the DAGS
energy. We first note that the rhs of Eq.(16) for any given realization of
$W(x)$
can be bounded from above

\[\frac{1}{\int^{0}_{-R} [\varphi_{0}^{(1)}(x)]^{- 2} \; dx \;
\int^{R}_{- R} [\varphi_{0}^{(1)}(x')]^{2} \; dx'} \; +  \quad\]
\[ \; + \;  \frac{1}{\int^{R}_{0} [\varphi_{0}^{(1)}(x)]^{- 2} \; dx \;
\int^{R}_{- R} [\varphi_{0}^{(1)}(x')]^{2} \; dx'}  \; \leq  \quad\]
\[\leq \; \frac{1}{\int^{0}_{-R} [\varphi_{0}^{(1)}(x)]^{- 2} \; dx \;
\int^{0}_{- R} [\varphi_{0}^{(1)}(x')]^{2} \; dx'} \; +  \quad\]
\[ \; + \;  \frac{1}{\int^{R}_{0} [\varphi_{0}^{(1)}(x)]^{- 2} \; dx \;
\int^{R}_{0} [\varphi_{0}^{(1)}(x')]^{2} \; dx'}  \; = \quad\]
\[ = \; \frac{1}{\int^{0}_{-R}  \int^{0}_{-R} dx \; dx'
\; \exp( 2 W'(x') - 2 W'(x))} \; + \quad\]
\[ \; + \; \frac{1}{\int^{R}_{0}  \int^{R}_{0} dx \; dx'
\; \exp( 2 W(x') - 2 W(x))}  \quad (49)\]
As one may readily notice, the inequality in Eq.(49) is obtained by simply
 diminishing the
limits of integration; in the first term we change the limits of integration
over the
variable $x'$ from $[-R,R]$ to $[-R,0]$, while in the second one the limits
are changed from
$[-R,R]$ to $[0,R]$.
Since $\varphi_{0}^{(1)}(x)$ is positive
definite, the diminishing of limits decreases the value of the integral
and consequently, increases the terms on the rhs of Eq.(49).

Now we will  try to
 find an appropriate functional
of the extremes of the random function $W(x)$ which will bound
 the integrals in Eq.(49)  from below, and thus in such a way will enhance
 the bound in Eq.(49).

We note here parenthetically
that this problem turns to be rather non-trivial.
 In particular, standard integral
inequalities (such as, for instance, the Schwartz inequality) are obviously
insufficient since
they predict an algebraic growth of the integral
\[\int^{\pm R}_{0}  \int^{\pm R}_{0} dx \; dx'
\; \exp( 2 W(x') - 2 W(x)), \quad (50)\]
while the simple analysis of the "typical" behavior shows that Eq.(50) grows
at least as a  stretched-exponential function of $R$. In addition, the
integrands in Eq.(50) don't possess well-defined
derivatives and thus one can not expand the integrands in the vicinity of the
extremes of function $W(x)$ and make use
of the standard saddle-point-like estimates.

To illustrate the derivation  of such a bound we first turn to more lucid
discrete-space
picture, assuming
that $x$ and $y$ are discrete variables $x,y = 0,1, ... ,R$, and then
approximate  the integrals in Eq.(50) as products of two sums

\[\int^{\pm R}_{0} dx \int^{\pm R}_{0}  dx'
\; \exp( 2 W(x') - 2 W(x)) \; \approx \;
\sum_{x = 0}^{R} \sum_{y = 0}^{R} \; a_{xy} \quad (51.a)\]
with
\[ a_{xy} \; =
\; \exp(2 W(x) - 2 W(y)) \quad (51.b)\]
The derivation of
 the corresponding upper bound in the continuous-space, which is substantially
more
 lengthy, will be
merely outlined  in the Appendix to this paper.

We notice that the rhs of Eqs.(51) is the sequence of $(R + 1)^{2}$ positive
terms, each of which is an exponential of the distance between the positions
of a
given trajectory $W(x)$ taken  at two different moments of "time" $x$ (summed
over
 all
possible $x$ from the interval $[0,R]$).
 From this sequence of positive terms $\{a_{xy}\}$ we choose the maximal term,
$max_{x,y \epsilon [0,R]}\{a_{xy}\}$,
which is
evidently the
exponential of the difference of the maximal positive displacement $M_{+}$,
($M_{+} >
0$),
of the trajectory
 $W(x)$
(which is achieved at some moment $x = x^{*}$)
and the maximal negative displacement $M_{-}$, ($M_{-}  < 0$), of the same
trajectory
 (achieved
at the moment $y = y^{*}$, both $x^{*}$ and $y^{*}$ belonging
to the interval $[0,R]$),
\[M_{+} \; = \; max_{x \epsilon  [0,R]}\{W(x)\} \; = \; W(x^*)      \quad
(52)\]
\[M_{-} \; = \; min_{x \epsilon  [0,R]}\{W(x)\}  \; = \; W(y^*)    \quad (53)\]
Since all $a_{xy} \geq 0$, the sum in the rhs of Eq.(51) is evidently
larger than the maximal term
 of this sequence, i.e.

\[\sum_{x = 0}^{R} \sum_{y = 0}^{R} \; a_{xy} \; \geq \; max_{x,y \epsilon
[0,R]}\{a_{xy}\} \; = \;
exp(2 (M_{+} - M_{-})) \quad (54)\]
Eq.(54) represents  the (discrete-space) formulation of the
desired bound on the integrals in
Eq.(50).

Let us now see how this bound can be employed for the derivation of the upper
bound on
the DAGS. Making use of Eqs.(49) and (54) we have that, at a given realization
of
$W(x)$, the
ground-state energy can be bounded from above as

\[E_{0}(R, \{\phi\}) \; \leq \; \exp( - 2 S') \; + \; \exp( - 2 S),  \quad
(55)\]
where we denote by

\[ S' = max_{x \epsilon [0, - R]}\{W'(x)\} - min_{x \epsilon [0, - R]}\{W'(x)\}
\; = \; M_{+}' \; - \; M_{-}'  \quad\]
 and
\[S = max_{x \epsilon [0, R]}\{W(x)\} - min_{x \epsilon [0, R]}\{W(x)\} \; = \;
M_{+} \; - \; M_{-}  \quad\]
Random variables as $S'$ (or $S$) are
known in the statistics of random walks as a span of random walk (see Fig.3),
which can be
visualized (in d-dimensional space)
as the dimensions of the smallest box with  sides parallel to the coordinate
axes that
entirely contain the trajectory of a random walk \cite{weiss}.
The exact probability
distribution $P(S,R)$ of random variable $S$  is well-known \cite{weiss}; in
the case of large $R$, a convenient
representation reads

\[P(S,R) \; = \; \frac{8 \sigma R}{S^{3}} \; \sum^{\infty}_{k = 0}
[\frac{\pi^{2}
(2k + 1)^{2} \sigma R}{S^{2}} - 1] \; \exp( - \frac{\pi^{2} (2k + 1)^{2} \sigma
R}{2 S^{2}}) \quad (56)\]

Therefore, the calculation of the upper bound on the
DAGS energy reduces to the calculation of the integral

\[E_{0}(R) \; \leq \; 2 \int^{\infty}_{0} dS \; \exp( - 2 S) \; P(S,R)  \quad
(57)\]
in which, noticing that $S$ and $S'$ have identic distribution
 functions
(although
 for given realizations of $W(x)$ and $W'(x)$ they may have
different values) the contribution
of $S'$ may be simply accounted by introducing a multiplier $2$.

Let us now consider the behavior of the integral in Eq.(57) in the limit of
large $R$.
We first note that in this limit in Eq.(56) only the
 term with $k = 0$
is relevant. Second, noticing  that the integrand is a bell-shaped function, we
perform
 the integral
using the saddle-point approximation. Maximizing the terms in the exponent we
 get that the
saddle-point depends on $R$ as

\[S \; = \; S^{*} \; = \; (\frac{\pi^{2} \sigma R}{2})^{1/3} \quad (58)\]
and consequently, the bound in Eq.(57) attains the form

\[E_{0}(R) \; \leq \; 32 (\frac{2 \sigma R}{3 \pi})^{1/2}
\; \exp( - \frac{3}{2^{1/3}} \;
( \pi^{2} \sigma R)^{1/3}) \quad (59)\]

Therefore, the upper bound on the DAGS energy shows a
 stretched-exponential dependence on $R$
with the characteristic exponent $z = 1/3$, i.e. aside
 from the numerical factor $2^{-1/3}$
in the exponent and pre-exponential multipliers (which
 are not reliable in view of the
approximation involved), essentially the same behavior as
the lower bound in Eq.(48). Since
both lower and upper bounds have the same dependence on $R$
 and also, on physical grounds,
$E_{0}(R)$ is a monotonically decreasing function of $R$, we may infer that the
stretched-exponential dependence with $z = 1/3$ is the asymptotically exact
 result for $E_{0}(R)$.
It is also important to note that both the lower and the upper bounds turn out
to be
 supported
by the same "class" of trajectories $W(x)$, such that $W(x) \propto x^{1/3}$.

To close this subsection we remark that the coincidence in the $R$-dependence
of
 the
lower and the upper bounds is, in essence, due  to the fact that the measure of
the
restricted trajectories, used in the derivation of the lower bound, and the
probability
distribution of the maximal displacement (or of the span $S$)
 are intrinsically related to
each other \cite{weiss}. Actually,
 the probability $P(A,R)$ that a random walker,
starting at the moment $R =
0$ at the origin, remains within an interval $[ - A, A ]$ in an $R$-step walk
is
 just
the probability that the maximal displacement of this random walker is less
than
 $A$.
It is clear that the probability of having the maximal displacement exactly
equal to
$A$ is given by \cite{weiss}

\[V(A,R) \; \approx \; \frac{\partial P(A,R)}{\partial A}, \quad (60)\]
and consequently, the probability of having the span of an $R$-step walk equal
to $S$
will follow

\[P(S,R) \; \approx \; \frac{\partial P(A,R)}{\partial A}_{| A = S/2}, \quad
(61)\]

Further on, evaluating the lower bound we have searched for such an $A$ which
maximizes the product $\exp( - 4 A)
 P(A,R)$.
On the other hand,  the upper bound was eventually reduced to
 the integral in
Eq.(57), which, using Eq.(61) can be rewritten as

\[\int^{\infty}_{0} dS \; \exp( - 2 S) \; P(S,R) \; \approx \;
\int^{\infty}_{0} dS \; \exp( - 2 S) \; \frac{\partial P(A,R)}{\partial A}_{| A
= S/2}
\quad (62)\]
Integrating Eq.(62) by parts we arrive at performing an integral with the
integrand
$\exp( - 4 A)
 P(A,R)$.
Since the integrand is a bell-shaped function of $A$ and thus
the saddle-point approximation can be used, calculation of this integral also
reduces
to maximizing the integrand.

\subsection{Random Gaussian potentials $\phi(x)$ with correlations.}

In this subsection we will briefly discuss the behavior of the DAGS in the case
where
fluctuations of $\phi(x)$ are correlated.

Now, as we have already mentioned
the relevant property is the integral $W(x) = \int^{x}_{0} dx' \; \phi(x')$,
rather than $\phi(x)$ itself.  It is therefore convenient to define
the correlations in random potential $\phi(x)$ in terms of
the integral $W(x)$. We consider here the case where $W(x)$ is zero in average,
as
 in Eq.(2), and define the second moment as follows

\[< W(x) W(x') > \; \sim \; |x - x'|^{1 + \lambda},  \; - 1 \; \leq \; \lambda
\;
\leq \; 1    \quad (63)\]
The parameter $\lambda$ in Eq.(63) determines the nature of correlations in the
random
potential $\phi(x)$. The border-line
case $\lambda = 0$ corresponds to delta-correlated fluctuations of $\phi(x)$,
when
$W(x)$ is a trajectory of the conventional Brownian motion.
This case has been examined in detail
in previous sections.
The case of positive $\lambda$, ($\lambda > 0$), describes the situations
 in which fluctuations of $\phi(x)$ in two neighboring points $x$ and $x'$
tend to be of the same sign. Here the trajectories $W(x)$ have strong
persistency;
thinking in terms of random walk one may say that here the random walker most
likely
continues the motion in the  direction of the previous step than changes
 the direction of  motion. Consequently, its trajectories
are more "swollen" and spatially more extended compared to the case $\lambda =
0$.
Finally, the case $\lambda < 0$ describes disorder with negative correlations
when the
values of the potential $\phi(x)$ in two neighboring points $x$ and $x'$ tend
to
 have
different signs. Here the random walker has a tendency of changing the
direction of its motion at each step and its
trajectories $W(x)$ are essentially more compact, compared to
the case of conventional random walk.

Using Eq.(63) one can readily estimate the typical behavior of the DAGS.
Since for the
typical realizations of $W(x)$ one expects that $W(x) \sim x^{(1 +
\lambda)/2}$,
 we
will have $ < ln \tau_{+}(0,R) > \sim R^{(1 + \lambda)/2}$, and consequently,

\[E_{0}(R)_{typ} \; \sim \; exp( - R^{(1 + \lambda)/2}) \quad (64)\]

Consider now the behavior of the DAGS stemming from atypical realizations and
generalize the formalism employed for the derivation of the lower bound.
Anticipating
the reasonings which underly the inequality in Eq.(36) and Eqs.(41), we have
that the
DAGS can be estimated as

\[E_{0}(R) \; \geq \; \exp( - 4 A) \; P_{\lambda}^{2}(A,R), \quad (65)\]
where $P_{\lambda}(A,R)$ denotes the probability that a random walker, which is
at the
origin at $R = 0$ and whose trajectories obey Eq.(74) will remain inside the
strip
  $[- A, A]$ during the time interval  $[0, R]$. Such a probability can be
estimated as
\cite{weiss,blu}:

\[P_{\lambda}(A,R) \; \sim \; \exp( - R/ A^{d_{\omega}}),   \quad (66)\]
where $d_{\omega} = 2/(1 + \lambda)$ is the "fractal" dimension \cite{blu}
of the random walk
defined by Eq.(63).  Plugging Eq.(66) into the Eq.(65) and maximizing the
product
with respect to $A$ we obtain the following estimate

\[E_{0}(R) \; \sim \; \exp( - R^{1/(1 + d_{\omega})}), \; R \; \gg \; 1, \quad
(67)\]
or, in terms of the parameter $\lambda$,

\[E_{0}(R) \; \sim \; \exp( - R^{(1 + \lambda)/(3 + \lambda)}), \; R \; \gg \;
1
 \quad (68)\]

Behavior as in Eqs.(67) and (68) is thus supported
 by such atypical trajectories $W(x)$ which
grow with $x$ as $x^{(1 + \lambda)/(3 + \lambda)}$. It is important to note
that
 again
the estimate in  Eqs.(67) and (68)  shows a slower dependence on $R$
as compared to the typical
behavior in Eq.(64).

To close this section we explain what we have in mind when
saying that realizations of
disorder which support the anomalous stretched-exponential behavior of the DAGS
share
common features with the realizations of trajectories which support the
anomalous
long-time decay of the survival probability of a particle diffusing in the
presence of
randomly placed traps or Lifschitz tails in the low energy density of states of
an
electron in the presence of randomly dispersed scatterers \cite{bal,rya}.

Let us remind, on the example of the trapping problem, some basic formulations
and
results. Suppose a one-dimensional, infinite in both directions, line with
 immobile
traps $B$ which are placed completely at random at mean concentration $n_{B}$.
 At $t = 0$ we introduce on the line
some concentration of particles of another type, say $A$, and let them diffuse
independently of each other. As soon as a $A$ particle approaches a $B$ trap,
the $A$ particle gets annihilated,
while the trap is
unaffected. The question of interest is to define the time evolution of the
concentration of $A$ particles (or the survival probability), averaged with
respect to
the spatial arrangement of traps.

Let $C(x,t)$ denote the local concentration of $A$ particles at the point $x$
at
 time
$t$. It obeys the diffusion equation

\[\dot{C}(x,t) \; = \; D \; \frac{\partial^{2}}{\partial x^{2}} C(x,t), \quad
(69)\]
where $D$ is the diffusion coefficient of $A$ particles. Eq.(69)
is to be solved subject to the adsorbing boundary conditions imposed at the
points occupied by traps; that is

\[C(x = X_{i},t) \; = \; 0,  \quad (70)\]
for any $X_{i}$ from $\{X_{i}\}$, where $X_{i}$ defines the position of the
$i$-
th
trap, $ - \infty \leq i \leq \infty$, and $\{X_{i}\}$ denotes the set of traps'
positions.

A nice feature of the one-dimensional geometry is that this problem can be
solved
exactly \cite{bal,weissa}, by simply noticing that evolution of $C(x,t)$
on some interval $[X_{i},X_{i+1}]$ is independent of other intervals.
Consequently, one has to find the solution of
 the diffusion equation on a finite
interval of fixed length $W$, subjected to the adsorbing boundary conditions at
the
ends of the interval, and then perform averaging with respect
to the distribution of the interval's
length. Such a solution is given by Eq.(43), which in the limit of sufficiently
large
times reads
\[P(W,t) \; \approx \;  \exp \left( -  \pi^2 \frac{D t}{W^{2}}\right)  \quad
(71)\]
Now, the disorder-average concentration of $A$ particles at time $t$ will be
defined as

\[<C(x,t)> \; \approx \;  \int^{\infty}_{0} dW \; P(W,t) \; P(W), \quad (72)\]
where $P(W)$ is the probability of having a trap-free interval of length $W$.
For
Poisson distribution of traps $P(W)$ behaves as

\[P(W) \; \propto \; \exp( - n_{B} W) \quad (73)\]
Substituting Eqs.(73), (71) into the Eq.(72) we thus arrive at an integral of
essentially the same structure as that in Eq.(57), which yields
\cite{bal,don,weiss,blu,weissa}:

\[<C(x,t)> \; \approx \;  \exp\left( - 3\left({\pi^2 \over 4}n_B^2
D t)^{1/3}\right)\right) \quad (74)\]

The behavior as in Eq.(74) shows that the long-time decay of the
disorder-average
concentration is supported by such bounded
realizations $W(t)$ of $A$ particles' random walks which obey $|W(t)| \leq A
\propto
t^{1/3}$, i.e. the same class of trajectories which support the
large-$R$ behavior of the DAGS in the problem studied in the present paper.

\section{Conclusions.}

To conclude, we have studied a new aspect of a one-dimensional localization
problem
associated with the supersymmetric Hamiltonian in Eq.(1) in which the potential
$\phi(x)$ is Gaussian random function of the spatial variable $x$. We have
derived an
explicit expression for the ground-state energy of the Hamiltonian (1) defined
on a finite interval  of the $X$-axis for a given realization
of disorder and analysed the
dependence of the disorder-average ground-state energy on the length
 $R$ of the interval. We
have shown that it is described by a stretched-exponential function of the form
$\exp( - R^{z})$, in which the characteristic exponent $z$ is dependent merely
 on the nature
of correlations in random potential.
In case when fluctuations in random potential are
delta-correlated we found $z = 1/3$.
In case when fluctuations are defined by Eq.(74)  we have deduced that $z = (1
+
\lambda)/(3 + \lambda)$.
We have shown that such a behavior is quite different from the one expected
when
only typical realization of disorder are considered and thus is
supported by atypical realizations of random potential, which behave as

\[ \int^{x} dx' \; \phi(x') \; \propto \; x^{(1 +
\lambda)/(3 + \lambda)} \quad \]
We have also shown that such realizations
belong to the  class  of trajectories which support an anomalous
long-time behavior of the survival probability of a random walk in the presence
of
randomly placed traps.

\begin{center}
\begin{Large}
Appendix
\end{Large}
\end{center}

In this appendix, we outline the derivation of the upper-bound for the DAGS
in the continuous-space limit.
 Consider the integral in Eq.(50) (for
simplicity we suppose that limits of the integration are from $0$ to $ + R$)
and, as it
was done before, assume that a given trajectory $W(x)$ reaches its maximal
value
 at the
point $x = x^{*}$ and its minimal value - at the point $x = y^{*}$. Let us
choose some
positive constant $\varepsilon$, such that $0 < \varepsilon \ll R$ and $x^{*} +
\varepsilon, y^{*} +
\varepsilon  \leq R$.
Since the integrand in Eq.(50)
is positive definite, the following inequality holds

\[\int^{R}_{0} dx \int^{R}_{0} dy \; \exp(2 W(x) - 2 W(y)) \; \geq \;
\int^{x^{*} + \varepsilon}_{x^{*}} dx \int^{y^{*} + \varepsilon}_{y^{*}} dy
\; \exp(2 W(x) - 2 W(y))   \quad (A1)\]
Now, taking advantage of the inequality in Eq.(49) we have for the DAGS:

\[E_{0}(R) \; \leq \; <\frac{1}{\int^{R}_{0} dx \int^{R}_{0} dy \; \exp(2 W(x)
-
 2
W(y))}> \; + \; \quad\]
\[ + \; <\frac{1}{\int^{0}_{-R} dx \int^{0}_{-R} dy \; \exp(2 W(x) - 2
W(y))}> \; =  \quad\]
\[ =  \; 2 \; <\frac{1}{\int^{R}_{0} dx \int^{R}_{0} dy \; \exp(2 W(x) - 2
W(y))}> \; = \; \quad \]
\[ = \; 2 \; \int^{\infty}_{0} dS \; P(S,R) \; \int \int dM_{+} \; dM_{-} \;
\delta(S -
M_{+} + M_{-}) \; \times \quad\]
\[ \times \;  <\frac{1}{\int^{R}_{0} dx \int^{R}_{0} dy \; \exp(2 W(x) - 2
W(y))}>_{|(M_{+} = W(x^{*}); M_{-} = W(y^{*}))}, \quad (A2)\]
where the brackets with the subscript $(M_{+} = W(x^{*}); M_{-} = W(y^{*}))$
mean that
the average is taken with respect to the trajectories $W(x)$ whose maximal
positive displacement is equal to $M_{+}$ and the maximal negative displacement
is
equal to $M_{-}$.

Further on, the inequality in Eq.(60) enables us to enhance the bound in
Eq.(61)
 and
write

\[E_{0}(R) \; \leq  \; 2 \; \int^{\infty}_{0} dS \; P(S,R) \; \int \int dM_{+}
\; dM_{-} \; \delta(S - M_{+} + M_{-}) \; \times \quad\]
\[ \times \; <\frac{1}{\int^{x^{*} +
\varepsilon}_{x^{*}} dx \int^{y^{*} + \varepsilon}_{y^{*}} dy \; \exp(2 W(x) -
2
W(y))}>_{|(M_{+} = W(x^{*}); M_{-} = W(y^{*}))} \quad (A3)\]

Let us now estimate the value of the following functional:

\[E \; = \; \int^{\infty}_{0} dS \; P(S,R) \; \int \int dM_{+} \; dM_{-} \;
\delta(S -
M_{+} + M_{-}) \; \times \quad\]
\[ \times \; <\frac{1}{\int^{x^{*} +
\varepsilon}_{x^{*}} dx \int^{y^{*} + \varepsilon}_{y^{*}} dy \; \exp(2 W(x) -
2
W(y))}>_{|(M_{+} = W(x^{*}); M_{-} = W(y^{*}))} \quad (A4)\]
To do this we enclose the points $W(x^{*})$  and $W(y^{*})$ by circles of
radius
$\delta$ (see Fig.3), where $\delta = \delta(R)$ is a slowly growing function.
The choice of the dependence $\delta(R)$ will be made later.
Further on, we  divide the set of all possible trajectories $\Omega$ into two
different subsets. The first subset $\{A\}$ comprises all such trajectories
$W(x)$ of
random walk (with its maxima at $M_{+}$ and minima at $M_{-}$) which, on the
interval $x \; \epsilon \;
[x^{*},x^{*} + \varepsilon]$ don't cross the circle around the point $W(x^{*})$
and on the interval $x \; \epsilon \;  [y^{*},y^{*} + \varepsilon]$ don't
cross the circle
around $W(y^{*})$ (e.g. the trajectory $1$ in Fig.3).
The subset $\{B\}$ comprises the rest of the trajectories (for instance, the
trajectory $2$ in Fig.3).
We write now

\[E \; = \; A \; + \; B, \quad (A5)\]
where $A$ stands for the average of the integrand in Eq.(62) with the
trajectories
forming the subset $\{A\}$, while $B$ denotes the contribution to $E$ coming
from the
average of the integrand over of the trajectories forming the subset $B$.

Consider first the contribution coming from the trajectories in the subset
$\{A\}$.
By definition of $\{A\}$, we have that on the interval  $[x^{*},x^{*} +
\varepsilon]$
the trajectory $W(x)$ obeys the inequality $M_{+} - \delta < W(x) \leq M_{+}$;
and on
the interval $[y^{*},y^{*} + \varepsilon]$ the trajectory $W(y)$
 obeys $M_{-} \leq W(y) < M_{-}  + \delta$.
Consequently, the integrands in Eq.(A4) are bounded from below by

\[\exp(2W(x)) \; \geq  \; \exp(2 M_{+} - 2 \delta), \quad (A6)\]
\[\exp(- 2 W(y)) \; \geq  \; \exp(- 2 M_{-} - 2 \delta), \quad (A7)\]
and thus $A$ is majorized by

\[A \; \leq \; \frac{\exp(4 \delta)}{\varepsilon^{2}}
\; \int^{\infty}_{0} dS \; P(S,R) \; \exp( - 2 S)   \quad (A8)\]
Next we estimate the contribution from the trajectories of the subset $\{B\}$.
Here,
for $x \; \epsilon \;
[x^{*},x^{*} + \varepsilon]$ and $y \; \epsilon \; [y^{*},y^{*} + \varepsilon]$
the function $\exp(W(x) - W(y))$ is always greater than $1$ and consequently

\[\int^{x^{*} +
\varepsilon}_{x^{*}} dx \int^{y^{*} + \varepsilon}_{y^{*}} dy \; \exp(2 W(x) -
2
W(y)) \; \geq \; \varepsilon^{2} \quad (A9)\]
Accordingly, the contribution coming from the trajectories of the subset
$\{B\}$
 can
be majorized by

\[B \; \leq \; \frac{1}{\varepsilon^{2}} \; P(\{B\}), \quad (A10)\]
where $P(\{B\})$ denotes the measure of trajectories forming the subset $B$.
When
$\delta$ is chosen such that
$\delta^{2} \gg 2 \sigma \varepsilon$, this measure vanishes as

\[ln P(\{B\}) \; \sim \; - \frac{\delta^{2}}{2 \sigma \varepsilon}  \quad
(A11)\]

Now, gathering Eqs.(A8) and (A10) we have that $E$ is bounded from above by

\[E \; \leq \; \frac{\exp(4 \delta)}{\varepsilon^{2}}
\; \int^{\infty}_{0} dS \; P(S,R) \; \exp( - 2 S) \; + \;
\frac{1}{\varepsilon^{2}} \;
\exp(- \frac{\delta^{2}}{2 \sigma \varepsilon}) \quad (A12)\]
Our previous analysis shows that the integral over the span variable $S$ in the
first
term on the rhs of Eq.(A12) vanishes with $R$ as a stretched-exponential
function of the
form $\exp(- R^{1/3})$. Thus, the rhs of Eq.(A12) behaves as

\[\approx \; \frac{\exp(4 \delta)}{\varepsilon^{2}}
\;  \exp( - R^{1/3}) \; + \; \frac{1}{\varepsilon^{2}} \;
\exp(- \frac{\delta^{2}}{2 \sigma \varepsilon}) \quad (A13)\]
Now we have to make the choice of $\varepsilon$ and $\delta(R)$. One readily
notices
that the proper choice will be if we suppose that $\varepsilon = constant$ and
$\delta(R) \sim R^{\gamma}$, where $\gamma$ is an arbitrary number from the
interval
$]1/6,1/3[$. If $\gamma > 1/6$ the second term on the rhs of Eq.(A13)
 vanishes with $R$ faster than the first term and thus
the leading large-$R$ behavior will be given by the first
term on the rhs of Eq.(A13). On the other hand
the requirement $\gamma < 1/3$ insures that the
leading large-$R$ behavior follows the $\exp(- R^{1/3})$
dependence, since $R^{-1/3} \delta(R)
\rightarrow 0$ when $R \rightarrow \infty$.

Therefore, we have shown that also in the continuous-space limit the upper
bound
 on the
DAGS vanishes with $R$ as a stretched-exponential function with the
characteristic
exponent $z = 1/3$. The bound derived here (although it suffices to prove the
asymptotically exact dependence $\exp(- R^{1/3})$) turns out, however, to be
worse
 than the
one found in the discrete-space case; it differs from the bound in Eq.(59) by
an
additional multiplier which grows with $R$ as $\exp(R^{\gamma})$. Besides, this
bound
is not optimal; there are no well-defined values of $\varepsilon$ and $\delta$
which
minimize the upper bound. Apparently, an optimal upper bound in the
continuous-space
can be also devised, but this is beyond the aims of the present paper.

\begin{center}
\begin{Large}
Acknowledgments.
\end{Large}
\end{center}

G.O. acknowledges the hospitality and
the financial support from  Division de Physique Th\'eorique of the IPN, Orsay,
 and CNRS.
S.F.B. is supported by the ONR  Grant N 00014-94-1-0647.

\pagebreak

\begin{center}
\begin{Large}
Figure Captions
\end{Large}
\end{center}

Fig.1a. Schematic picture of a two-level random potential $\phi(x)$.

Fig.1b. An auxiliary configuration $\tilde{\phi}(x)$.

Fig.2.  Schematic representation of the sets $\Omega$, $\Omega'$ and the
subsets
$\omega$, $\omega'$. The set $\Omega$ (dashed triangle on the half-plane $x >
0$) comprises all possible realizations of
an $R$-step random walk trajectories $W(x)$ with $x \epsilon [0,R]$.
 The set $\Omega'$
(dashed triangle on the half-plane $x <
0$) comprises respectively all possible trajectories of an $R$-step random walk
with $x \epsilon [0,- R]$. The subsets $\omega$ and $\omega'$ are the areas cut
from
the sets $\Omega$ and $\Omega'$ by the lines $W(x) = A$ and $W(x) = - A$.

Fig.3.  Maximal positive, maximal negative displacements and the span
of the trajectory $W(x)$ with $x$ defined on
the interval $[0,R]$.

\end{document}